\let\new=\newcommand
\new{\be}{\begin{equation}}
\new{\ee}{\end{equation}}
\new{\bea}{\begin{eqnarray}}
\new{\eea}{\end{eqnarray}}
\new{\bdm}{\begin{displaymath}}
\new{\edm}{\end{displaymath}}
\new{\B}{{\bf B}}
\new{\je}{{\bf j}}
\new{\vel}{{\bf v}}
\new{\fv}{{\bf f}}
\new{\ov}{{\bf \omega}}
\new{\curl}{ \nabla \times}
\new{\Div}{ \nabla \cdot}
\new{\lap}{\nabla^2}
\new{\del}{\partial}
\new{\nor}{\nabla \chi}
\new{\dt}{\rm{d}_t}
\new{\vchi}{{\rm v}_\chi}
\new{\E}{{\bf E}}
\new{\rv}{{\bf r}}
\new{\vb}{{\bf b}}
\new{\av}{{\bf a}}
\new{\cv}{{\bf c}}
\documentstyle[]{article}
\def\thebibliography#1{\small \section*{\s References}
\list
 {\labelenumi}{\setlength\labelwidth{1.4em}\leftmargin\labelwidth
 \setlength\parsep{0pt}\setlength\itemsep{0pt}
 \setlength{\itemindent}{-\leftmargin}
 \usecounter{enumi}}
 \def\newblock{\hskip .11em plus .33em minus -.07em}
 \sloppy
 \sfcode`\.=1000\relax}

\font\t=cmbx10 at 16pt
\font\au=cmr10
\font\ad=cmti10 at 9pt
\font\s=cmbx10
\font\sb=cmti10

\def\pmb#1{\setbox0=\hbox{$#1$}%
\kern-.025em\copy0\kern-\wd0
\kern.05em\copy0\kern-\wd0
\kern-.025em\raise.0433em\box0}
\begin{document}
\begin{center}
{\t A unified description of anti-dynamo conditions for incompressible flows} 
\end{center}
\vspace{0.7cm}
\begin{center}
{\au A. Mangalam}\\ {\ad Indian Institute of Astrophysics,
\\ Koramangala \\
Bangalore 560034, INDIA \\
Internet: mangalam@iiap.res.in}
\end{center}
\vspace{1 cm}

\hrule
\vspace{0.2 cm}
A general type of mathematical argument is described, which applies to all the cases in which dynamo maintenance of a steady magnetic field by motion in a uniform density is known to be impossible.
Previous work has demonstrated that magnetic field decay is unavoidable under conditions of axisymmetry
and in spherical or planar  incompressible flows. These known results  are encompassed
by a  calculation  for flows described in terms of  a generalized poloidal-toroidal representation of the magnetic field with respect to an arbitrary two dimensional surface. We show that when  the velocity field is two dimensional, the dynamo growth, if any, that results, is linear in one of the projections of the field while the other projections  remain constant. We also obtain  criteria for the existence of and classification into  two and three dimensional velocity results  which are satisfied by a restricted set of geometries. In addition, we discuss the forms of spatial variation of the density and the resistivity that are allowed so that field decay still occurs for this set of geometries. 	
\vspace{0.2 cm}
\hrule
\vspace{0.5 cm}

{\au
\noindent
{\s 1. Introduction}
\\

\noindent
The steady state dynamo problem in a uniform density can be stated as follows. Given a uniform electrically conducting fluid, contained in a volume, what condition of must be satisfied by the velocity $\vel$ of a steady motion in order that a steady field $\B$ can be maintained by dynamo interaction between the motion and the field?
This classical problem of the magnetic dynamo concerns the question of the amplification or maintenance of the magnetic field in cases where the induction equation is valid.  The equations  yield only decaying solutions when the velocity and magnetic fields are both axisymmetric \cite{Cow34,BacCha56} or if the geometry has planar symmetry \cite{Zel56}. In both situations, the velocity and magnetic field  can be three dimensional but do not depend on at least one of the coordinates. In situations where the velocity is two dimensional, but the magnetic field is three dimensional,  the impossibility of dynamo action has been proven  if the flow is planar \cite{Mof78,Lor68,ZelRuz80} or spherical \cite{BulGel54,Bac58}. 

In the following, we use a generalized toroidal-poloidal representation of the magnetic field,
$\B$, with respect to an arbitrary two dimensional surface and derive two scalar equations 
for the poloidal and the toroidal potentials  from the induction  equation. We also prove a result, which is an extension of the one in \cite{RuzSok80},  that incompressible two-dimensional velocity  flows  in situations other than in the  above mentioned antidynamo theorems (where the field decays), lead to linear growth in one of the field components and are otherwise slow. Here, the fluid velocity is two dimensional in the sense that it lies entirely on surfaces which can be described by $\chi(\rv)=$ constant. The approach taken here also lends itself to a  unified and   simpler exposition of the previously cited results. In addition, we consider the special cases of spatially variable forms of the density and magnetic diffusivity. 

A closer study of the cases in which the dynamo maintenance of a steady field is impossible may throw light on
the general dynamo problem. The results suggest that the number if cases in which dynamo maintenance is impossible is restricted. We derive such criteria for the existence of  antidynamo result for a given geometry.
\\

\noindent
{\s 2. Normal  projection of the induction equation}
\\

\noindent
 The starting point of all the above investigations is the well known induction equation in MHD
\be
\del_t \B + \curl (\B \times \vel ) = - \curl (\eta \curl \B),
\label{indeq}
\ee
with the constraint
\be
\Div \B =0,
\label{sol}
\ee
  where $\eta = c^2/ 4 \pi \sigma$ is the magnetic viscosity and $\sigma$ is the conductivity.  As in the above cases, the fluid is assumed to be incompressible ($\Div \vel=0$).  We make the additional simplification of taking $\eta$ to be uniform. Later, we discuss the effects of relaxing the latter simplification.

The RHS of the equation (\ref{indeq})
represents resistive dissipation whereas the LHS  contains a term which represents
stirring of the $\B$ field by fluid motions. In analogy with the
heat conduction equation,  this shows that in a static fluid the fields decay, while stirring may induce field generation.  

Consider the fluid velocity and the magnetic field to be  described in terms of components perpendicular and parallel to the surfaces defined by $\chi(\rv)=$ constant. The quantities, $B_\chi \equiv \B \cdot \nor$ and $\vchi \equiv \vel \cdot \nabla \chi$, satisfy the reduced form of the induction equation projected normal to the surface 
\be 
\dt B_\chi - \B \cdot \nabla \vchi-\eta (\lap \B)_\chi =0,
\label{heat}
\ee 
where $\dt \equiv \del_t +\vel \cdot \nabla$, and the constraint (\ref{sol}) and  incompressibility condition was taken into account. One can  expand the third term of (\ref{heat}) using identities (\ref{a3})--(\ref{a5}), and  obtain
\be 
\dt B_\chi = \eta (\lap B_\chi - \Div [ (\nor \cdot \nabla) \B + (\B \cdot \nabla) \nor]) +\B \cdot \nabla \vchi,
\ee
where the last term, $\B \cdot \nabla \vchi$, equals zero when the velocity  fields lie on the surfaces, $\chi(\rv)=$ const.  Now  employing the identity (\ref{a4}) on the LHS, writing the first term on the RHS in terms of $\Div (B_\chi \nabla B_\chi)$  after multiplying throughout by $B_\chi$, and integrating, while taking $\vchi=0$ (we relax this later), this can be further reduced to the dissipation theorem \cite{RuzSok80},
\be
\frac{1}{2} \dt \int_V B_\chi^2 {\rm d^3}{ \rm \bf r} = -\eta \int_V  \{  (\nabla B_\chi)^2 +
B_\chi \Theta(\B,\chi)\} {\rm d}^3\rv,
\label{dissip}
\ee
where  a surface integral over $\Div (B_\chi \nabla B_\chi)$ obtained from Gauss's theorem vanishes at large distances.  Here we have introduced a useful quantity
\bea
\Theta({\bf Y}, \chi) &\equiv& \Div [ (\nor \cdot \nabla) {\bf Y} + ({\bf Y} \cdot \nabla) \nor] \nonumber \\
&=&  [ 2 \del_k (Y_i) + Y_k \del _i] \del_i \del_k \chi +\nabla \chi \cdot \nabla (\Div {\bf Y}).
\label{theta}
\eea
It can be seen that the last term in equation (\ref{theta}) vanishes if $\bf Y$ is solenoidal 
or if $\Div \bf Y$ is independent of a coordinate directed along $\nabla \chi$.  
The only positive contribution,  leading to growth, can come from the tensor term in $\Theta$ on the RHS of (\ref{theta}). Clearly,  if the surface  is planar ($\chi(\rv)= z$) or
 spherical ($\chi(\rv)= r^2/2$) the term becomes zero ($\del_i \del_k$ equals 0 or $\delta_{ik}$, respectively, and in the latter case the condition, (\ref{sol}) needs to be further applied),  implying that $B_\chi$ decays. This point was made by \cite{RuzSok80,ZelRuzSok83}. Hereafter, we suppress the notation $\Theta({\bf Y}, \chi)$ to $\Theta({\bf Y})$ unless $\chi$ is specified. In the following, where we keep the treatment general (by keeping the velocity and magnetic fields three dimensional), we find the same term occuring  in the surface projection of the induction equation. This enables us to expand upon these conditions for field decay and generalize the antidynamo results cited earlier. 
\\

\noindent 
{\s 3. Parallel projection of the induction equation}
\\

\noindent
It is convenient to express the magnetic field in terms of the local ``poloidal'' and ``toroidal'' components
\be
\B= \B_{\rm P} + \B_{\rm T}=\curl \curl (\psi \nor) + \curl ( \Phi \nor),
\ee 
where $\psi$ and $\Phi$ are the generalized poloidal and toroidal flux functions. This is analogous to the description of magnetic field given by \cite{Cha61,Mof78} for spherical geometry. We express the field in a coordinate system in which a special coordinate, $q$, is normal to the surfaces and is given by 
\be
\chi(\rv)= f(q).
\ee
The connection to the corresponding formulae in spherical geometry lies in the fact that any smooth surface has a local radius of curvature. Therefore, we implicity demand that the surface be smooth (have first derivatives defined). Next,  using properties (\ref{a7}, \ref{a2}) it can be seen that
\bea
B_\chi&=&\curl (\nabla
\psi \times \nor)\cdot \nor  \nonumber \\ &=& \Div [ (\nabla \psi \times
\nor) \times \nor]  \nonumber \\ &=& \Div [(\nabla \psi \cdot \nor) \nor
-(\nor)^2 \nabla \psi] \nonumber \\ &=& -(\nor \times \nabla)^2 \psi\equiv -\lap_\| \psi.
\label{b.n}
\eea 
 In order to examine the local components parallel to the surface, we take
a normal projection of the curl of the induction equation (\ref {indeq}),
\be
\del_t [(\curl \B) \cdot \nor)] +\nor \cdot [\curl \curl (\B \times \vel )] = -\nor \cdot [\curl \curl (\eta \curl \B)],
\label{normal}
\ee

After expanding $\B$, the
operand of the time derivative in the resulting equation  can be reduced in a fashion  similar to Eq.\ (\ref{b.n})
\be 
(\curl \B) \cdot \nor = -\lap_\| \Phi +\Theta(\curl [\psi \nor]).
\label{curlb.n}
\ee 
Defining, ${\cal C} \equiv \curl \B$, and using Eqs.\ (\ref{a3}) and (\ref{curlb.n}), the term on the RHS of
Eq. (\ref{normal}) yields,
\bea 
-\eta ({\rm curl} ^3 \B) \cdot \nor &=&  \eta (\lap {\cal C} )\cdot \nor
\nonumber \\ &=& \eta \left( \lap {\cal C}_\chi - \Div [ (\nor \cdot \nabla) {\cal C}
+ ( {\cal C} \cdot \nabla) \nor] \right ) \nonumber \\ &=&-\eta  \lap \lap_\| \Phi-
\eta \Theta({\cal C}) + \eta \lap \Theta(\curl (\psi \nor)),
\label{2diff}
\eea 
after performing manipulations identical  to those required in obtaining Eq. (\ref{theta}). The second term of equation (\ref{normal}) after applying Eq. (\ref{a7}), is
\be 
-\nor \cdot \{ \curl \curl [ \vel \times (\curl \curl (\psi \nor))  + \vel \times (\nabla \Phi \times \nor)] \}.
\label{exp}
\ee 
The second term may be evaluated in steps as follows
\bea
\vel \times (\nabla \Phi \times \nor) &=&\vchi \nabla \Phi- (\vel \cdot \nabla \Phi) \nor
\nonumber \\
\curl [\vel \times (\nabla \Phi \times \nor)]& = &\nabla \vchi \times \nabla \Phi-\nabla (\vel \cdot \nabla
\Phi) \times \nor \nonumber \\
-\nor \cdot \curl \{ \curl [\vel \times (\nabla
\Phi \times \nor)]\}& = &-\Div[\nabla \Phi(\nabla \chi \cdot \nabla \vchi)-\nabla \vchi(\nabla \chi \cdot \nabla \Phi)] \nonumber \\
&&- \lap_\|( \vel \cdot \nabla \Phi). 
\eea 
Defining, ${\cal D} \equiv \vel \times \B_{\rm P}$, ${\cal D}_\chi$ via (A.1) reduces to
\be
{\cal D}_\chi= \vel \cdot (\B_{\rm P} \times \nabla \chi).  
\label{dchi}
\ee
The first term in Eq. (\ref{exp}) will reduce to
\be 
-\nor \cdot \left [ \curl \curl {\cal D} \right ]= -\Theta( {\cal D})+\lap {\cal D}_\chi
\label{D}
\ee 
where the properties in the calculation (\ref{2diff}) are used. 
\\

\noindent
{\s 4. Results}
\\
 
Now  one can write the equation describing the evolution of the parallel components of the
field by including all the terms simplified to
the forms given in Eqs. (\ref{b.n}) --(\ref{dchi}), and rearranging terms, as
\bea
\lap_\| (\dt \Phi)  -\eta \lap \lap_\| \Phi-\eta \Theta({\cal C})& =&\Div[\nabla \vchi(\nabla \chi \cdot \nabla \Phi)-\nabla \Phi(\nabla \chi \cdot \nabla \vchi)]  \nonumber \\ &&
+(\partial_t-\eta \lap) \Theta(\curl[\psi \nor])-\Theta({\cal D}) + \lap {\cal D}_\chi.
\label{parallel}
\eea 
The full form of the corresponding equation for the normal component is 
\be
[\dt - \eta \lap] \lap_\| \psi -\eta \Theta(\B)=-(\B \cdot \nabla) \vchi,
\label{perp}
\ee
where the second term involves the normal component of velocity. 
\\

\noindent
{\sb 4.1 Non-diffusive flows $(\eta=0)$}
\\

\noindent
An exclusion theorem proved by \cite{RuzSok80} that states that two dimensional non-diffusive flows with a stationary velocity field and  the property $\vel \cdot \av=0$ (the flux helicity density) under  the gauge condition, $\Div  \av=0$, where $\av$ is  the vector potential for the velocity, will lead to a conservation of $\B \cdot \av$. We sketch their proof below. We write the Euler equation as
\be
\del_t \vel=\fv; ~~~~\fv=\nabla w -(\vel \cdot \nabla) \vel,
\ee
where all potential forces are collected in $w$. We can then write, under a gauge condition $\Div \av=0$,
\be
\dt (\av \cdot \vel)= \vel \cdot \nabla(\av \cdot \vel) + 2 \av \cdot \fv + \Div(\av \times 
{\rm curl}^{-1}\fv).
\ee
Next we substitute for $\fv$ and use
\be
-2 \av ( (\vel \cdot \nabla )\vel)= \Div (\av v^2)+2(\vel\times \ov)\cdot \av
\ee
 enroute to obtain
\be
\dt (\vel \cdot \av) = 2 \av \cdot(\vel \times \ov) + \Div [ (\vel \cdot \av) \vel + \av \times {\rm curl}^{-1}\fv +2(w-v^2/2) \av].
\ee
This implies that the flux helicity of the flow lines, $H_v \equiv \int \vel \cdot \av ~{\rm d}^3 x$, is conserved for Beltrami flows ($\vel \propto \ov $), or for potential flows ($\ov =0$). Similarly, in the flux freezing limit of the induction equation,
\be
\del_t \B = \curl (\vel \times \B); ~~~~~~\del_t \av= {\rm curl}^{-1} \fv
\ee
we can obtain the following after some transformations
\be
\dt(\av \cdot \B) =  \B \cdot \nabla (\vel \cdot \av) + \B \cdot {\rm curl}^{-1} \fv.
\ee
Now it is easy to see that for stationary flows ($\fv=0$), if the flux helicity, $H_v$, is zero everywhere,
that the cross helicity, $H_c\equiv \int \B \cdot \av ~{\rm d}^3 x$, is steady.

We now generalize this result by not imposing  any limitation on the  gauge  transformation-
as it is impossible to always simultaneously satisfy the conditions $\vel \cdot \av=0$ and $\Div \av=0$. When the velocity is two dimensional ($\vel_\chi=0$) and $\eta \rightarrow 0$,
it is clear from (\ref{perp}) that $B_\chi$ is an integral of motion and cannot grow with time. Further, if one takes a stationary velocity field with  $\vel \cdot \av=0$ and $\Div \av=0$, this implies that  the flow lines are along the intersection of two surfaces; i.e. $\vel= \nabla \xi \times \nabla \chi$. As a result, $B_\xi$ is another constant of motion ($\vel_\xi=0$). So the induction, $\vel \times \B= B_\xi \nabla \xi -B_\chi \nabla \chi$, is independent of time and the growth of the remaining component of $\B$, (along  \vel), can utmost be linear. Hence, we can conclude the intuitive result that  flows with zero linkage (and $\Div \av=0$) cannot lead to an exponential growth of the field, but utmost to a linear growth of the field in the direction of the flow.
\\

\noindent
{\sb 4.2 Antidynamo theorems}
\\

\noindent
Based on the  equations (\ref{perp}) and (\ref{parallel}), we can immediately divide the antidynamo results  for  two and three dimensional velocity fields. For the 3D results, we take both the velocity and the magnetic field to be three dimensional but invariant along the special coordinate q ($\del_q=0$). In the case of 2D results, the velocity field is two dimensional, $\vchi=0$, and the magnetic field is three dimensional. Under either of these conditions the first term on the RHS of (\ref{parallel}) vanishes.
The advection-diffusion operator ($\dt-\eta \lap$) can only manipulate the field and
cannot cause growth (cf. (\ref{dissip})). 

Here, an antidynamo case is defined as a situation in which the flux functions, $\psi$ and $\Phi \rightarrow 0$  everywhere (as $t \rightarrow \infty$) with the boundary conditions that these  flux functions vanish at remote surfaces which enclose the volume of fluid. So the strategy in finding antidynamo situations is to identify the conditions when one equation completely decouples from the other and the decay of the corresponding flux function kills the source terms in the other. It is natural to consider spherical and planar geometries first, since $\Theta(\B)$ and $\Theta( {\cal C})$ would then be zero (cf. (\ref{theta})). It is to be noted that $\Theta(\curl [\psi \nor])$ is zero for spherical, planar and axisymmetry.
\\

\noindent
{\sb 4.3 The cases of spherical and planar geometries}
\\

\noindent
For spherical geometry ($\chi= \frac{1}{2} r^2$) or planar geometry ($\chi=z$), the  equations (\ref{parallel})--(\ref{perp}) reduce to
\be 
[\dt -\eta \lap] \Phi  = {\cal Q}(\psi, \vel)
\label{ps}
\ee
\be
[\dt -\eta \lap] \lap_\| \psi = -\B \cdot \nabla \vchi
\label{sp}
\ee
where
\be 
{\cal Q}(\psi, \vel) \equiv {[\lap_\|]}^{-1}\{ \lap{\cal D}_\chi  -\Theta({\cal D}) -
\Div [\nabla \Phi (\nor \cdot \nabla \vchi) ]\},
\label{Q}
\ee 
and $\lap_\| $ represents $L^2$, the angular momentum operator in the spherical case or $\del_x^2 +\del^2_y$ in the planar case. In the two dimensional case ($\vchi=0$),
we have a source term only in the toroidal equation, (\ref{ps}), while $\psi$ decays. According to the definition of $\cal D$, as $\psi \rightarrow 0$, ${\cal D} \rightarrow 0$
 and the RHS of (\ref{Q}) vanishes.  Therefore ${\cal Q} \rightarrow 0$ and $\Phi$ will decay when $t \rightarrow \infty$. This follows from
the arguments after (\ref{dissip}), and  from potential theory
which  demands that the mean value of ${[\lap_\|]}^{-1} (0)$ is zero in the volume enclosed by a surface on which it vanishes. Physically, the normal component diffuses out and the field is confined to two dimensions; as a result, the field  is transported like a scalar, and hence decays. 

Now, in the three dimensional planar case ($\chi=z, \del_z=0$)  there is a source term only in the poloidal equation, (\ref{sp}).  It can be easily seen from (\ref{theta}) and (\ref{dchi}), that the toroidal source terms involving $\cal D$ and $\nor \cdot \nabla = \del_z$ are zero. Subsequently, as $\Phi \rightarrow 0$, $\B \cdot \nabla \rightarrow B_z \del_z (=0)$ and $\psi$ decays. It is interesting that in the three dimensional spherical  case none of the source terms in the above pair of equations are zero to begin with, and hence dynamo action occurs. 
\\

\noindent
{\sb 4.4 The case of axisymmetry}
\\

\noindent
  In axisymmetry ($\chi=\phi, \nor \cdot \nabla=\varpi^{-2}\del_\phi=0$) and $\cal D$ is poloidal.     As a result, ${\cal Q}=0$, as the term involving $\cal D$ is zero [cf. (\ref{D})], while $\Theta({\bf A}, \phi)= -(2 / \varpi) \del_\varpi (A_\phi/ \varpi)$, expressed in the cylindrical coordinates, ($\varpi, \phi,$ z). Then the equations, (\ref{parallel})--(\ref{perp}), simplify to
\be
D^2 [\dt -\eta D^2] \Phi =0,
\label{at}
\ee
\be
[\dt -\eta \varpi^{-2} D^2 \varpi^{2}] (B_\phi/\varpi)= - \B \cdot \nabla ({\rm v}_\phi/ \varpi),
\label{ap}
\ee
where $D^2 \equiv \lap-{2 \over \varpi} \del_\varpi$, and is known as the Stokes operator, and $\lap_\|$ represents $[\varpi^{-1}\hat{\phi} \times \nabla]^2=\varpi^{-2} D^2$.  Since the $D^2$ term can be reduced to a divergence term and subsequently to a surface integral by Gauss's theorem which vanishes due to the dipole behavior of the field at large radius, it does not contribute to growth; see for example, \cite{Mof78}(p.\ 114). Therefore $\Phi \rightarrow 0$, $\B \cdot \nabla \rightarrow B_\phi \del_\phi (=0)$ and $\psi$ decays according to similar arguments in \cite{Mof78} ( p.\ 115). 
\\

\noindent
{\sb 4.5 Effects of spatial variation of density and resistivity}
\\

\noindent
It is easy to see from the form of the continuity equation in steady state or under the anelastic approximation when the density is dependent only on the special coordinate q,
\be
\rho \Div \vel + {\rm v}_q \nabla_q \rho=0,
\ee
that $\Div \vel=0$ is  valid when $ {\rm v}_q=0$. Therefore only the above 2D results still hold. Non-uniform resistivity introduces a term $(\nabla \eta \times \nabla \chi) \cdot (\curl \B)$ in the equation (\ref{perp}) for $\psi$. This term is zero if $\eta$ is a function only of q and hence does not alter the decay of $\psi$.  This was commented upon by \cite{ZelRuz80} for the cases of spherical and planar geometries. We now show that $\Phi$ also decays in some special cases. The spatial variation of $\eta$ in q, however introduces a nonvanishing term $(\del_z \eta)(\del_z \Phi)$ for planar geometry or $(1/r)(\del_r \eta)(\del_r [\Phi r])$ for spherical geometry in equation (\ref{ps}). The 3D planar case follows trivially.  Now when $\vchi=0$, one can invoke a theorem on the resulting elliptic equation \cite{Vek62,Lor68}, which states that only the constant solution ($\Phi \equiv 0$) is possible under the condition of $\Phi$ vanishing at large distances. Similarly for axisymmetry, $\eta(\phi)$ introduces the term $(\del_\phi \eta){\cal C}_\varpi$ in (\ref{at}).  When there is no differential rotation, $B_\phi \rightarrow 0$ and ${\cal C}_\varpi$ which depends solely on $B_\phi$, vanishes and $\Phi$ decays as before. Therefore, the above 3D planar and 2D planar, spherical, and axisymmetric results are still valid if $\eta$ is a function only of q. Also, if there is no differential rotation, axisymmetric fields cannot be maintained if $\eta$ depends on $\phi$. It shown that in axisymmetry \cite{HidPal82}, poloidal fields cannot be maintained even by a compressible fluid with $\eta$ as a function of space and time. 
\\

\noindent
{\s 5. Concluding remarks}
\\

\noindent
In this paper we have cast the induction equation (\ref{parallel})--(\ref{perp}) in a geometry given by the surfaces $\chi(\rv)=$ constant. This was useful in extending a previous result for incompressible two dimensional flows while unifying, classifying, and simplifying the proofs of the known antidyanamo results, and thereby providing some new insights into the structure of the induction equation.  In order to deduce the general conditions  of decay, we can consider the order of decay of the flux functions, $\psi$ and $\Phi$. If $\Phi$ is to decay first, then all the potential source terms which are on the RHS of (\ref{parallel}) containing $\psi$ and $\cal D$ should be zero. This includes the condition that
\be
\lap {\cal D}_\chi -\Theta({\cal D})=0. 
\label{cond1} 
\ee
The above equation automatically ensures that $\Theta(\curl[\psi \nor])=\nor \cdot [\curl \curl \curl (\psi \nor)]$ is zero (c.f . (\ref{D})). The equation, (\ref{cond1}) is true for planar and axisymmetry. Consequently,  the 3D results of planar and axisymmetry follow, as the first term on the RHS of (\ref{parallel}) which involves $\nor \cdot \nabla$ is zero. On the other hand,  if $\psi$ is to decay first then the RHS of (\ref{perp}) should be zero demanding the 2D condition,  $\vchi=0$. Further, 
\be
\int_V [\Theta({\bf B}) -\lap {\bf B}_\chi ] {\rm d}^3 \rv \geq 0,
\label{cond2}
\ee
must hold  to ensure that $\psi$ decays as per (\ref{dissip}) and the  boundary conditions that the flux functions vanish at the remote boundaries.  This takes care of the decay of $\Phi$ since the remaining source terms,  which involve $\psi$, in (\ref{parallel}) vanish. The condition, (\ref{cond2}), is true for planar, spherical and axisymmetric geometries. The unifying aspect of treatment of the boundary conditions used here is that $[\lap_\|]^{-1}(0)$ is zero with vanishing flux at the boundaries.

We have been able show the impossibility of dynamo action (as defined in \S 4) with $\eta=$ const and $\Div \vel =0$, exists only for a restricted group of geometries as allowed by (\ref{cond1}) and (\ref{cond2}) for the case of 3D and 2D results respectively.  In addition,  the above results need further qualifications that $\vel$ is bounded for all time and that ``spiky''  time dependent behavior is excluded \cite{JamRobWin80}.  

As suggested in  a classic paper \cite{Cow57}, using a different approach, that the impossibility of a steady dynamo  for a given geometry  depends on the existence of an arbitrary current, $\bf j'$, such that $\int {\bf j} \cdot {\bf j'} {\rm d}^3\rv =0$, and that only a restricted set satisfies this equation.  Here we have derived specific conditions that determine these geometries. However,  a more rigorous analysis of the above two conditions is needed to find the  set of all possible $\chi$ that satisfies the above criteria.  In a paper, in preparation,
I investigate the geometries that  satisfy  the above criteria.
\\

\noindent
{\s Acknowledgment}: I thank P. H. Roberts for his valuable comments and for a critical reading of
the manuscript.
\\

\noindent
{\s Appendix: Formulae referenced in the text}
\\

\bea
\av \cdot (\vb \times \cv) &=& \vb \cdot (\cv \times \av) =\cv \cdot (\av \times \vb) \label{a1}\\
\av \times (\vb \times \cv)&=& \vb  (\cv \cdot \av) -\cv (\av \cdot \vb) \label{a2}\\
\curl (\curl \av) & = & \nabla (\Div \av) -\lap \av \label{a3}\\
\nabla( \av \cdot \vb) & = &(\av \cdot \nabla) \vb + (\vb \cdot \nabla) \av + \av \times (\curl \vb) + \vb \times (\curl \av) \label{a4}\\
\Div (\av \times \vb) & = &\vb \cdot ( \curl \av) - \av \cdot (\curl \vb) \label{a5}\\
\curl (\av \times \vb) & = & \av (\Div \vb) - \vb (\Div \av) + (\vb \cdot \nabla) \av - (\av \cdot \nabla) \vb \label{a6}\\
\nabla \{\times, \cdot \} (\psi \av) &=& \nabla \psi \{ \times, \cdot \} \av + \psi \nabla \{ \times, \cdot \} \av \label{a7}
\eea
}
\bibliography{unsrt}

\end{document}